\documentclass[aps,prl,twocolumn,groupedaddress]{revtex4}
\usepackage{amsmath}
\usepackage{graphicx}
\begin{document}

\title{Exciton Mott transition in Si revealed by terahertz spectroscopy}
\author{Takeshi Suzuki and Ryo Shimano}
\affiliation{Department of Physics, The University of Tokyo, Tokyo 113-0033, Japan}
\date{\today}

\begin{abstract}
We investigate the exciton Mott transition in Si by using optical pump and terahertz probe spectroscopy. The density-dependent exciton ionization ratio $\alpha$ is quantitatively evaluated from the analysis of dielectric function and conductivity spectra. The Mott density is clearly determined by the rapid increase in $\alpha$ as a function of electron-hole (e-h) pair density, which agrees well with the value expected from the random phase approximation theory. However, exciton is sustained in the high-density metallic region above the Mott density as manifested by the 1s-2p excitonic resonance that remains intact across the Mott density. Moreover, the charge carrier scattering rate is strongly enhanced slightly above the Mott density due to nonvanishing excitons, indicating the emergence of highly correlated metallic phase in the photo-excited e-h system. Concomitantly, the loss function spectra exhibit the signature of plasmon-exciton coupling, i.e., the existence of a new collective mode of charge density excitation combined with the excitonic polarization at the proximity of Mott density.

\end{abstract}

\maketitle

The electron-hole (e-h) system in photoexcited semiconductors has continued to attract interest over decades from the viewpoint of fundamental quantum many-body physics and also because of its crucial importance in semiconductor optoelectronic devices. As the e-h pair density increases, the charge neutral excitons, i.e., hydrogen-like e-h pairs that are usually formed in the low-density region, are dissociated into unbound electrons and holes and the system turns into a metallic e-h plasma phase. The transition from the insulating exciton gas to the metallic e-h plasma is referred to as the exciton Mott transition or crossover~\cite{Mott, Zimmerman1988, Zimmerman1978, Zimmerman1985}. Since the phenomenon is purely driven by electron-electron interaction, the phtoexcited e-h system offers an ideal playground to study the insulator-to-metal transition (IMT) in the ensemble of composite Fermionic particles. It is also crucially important to know to what extent excitons can exist and to understand how the two particle (e-h) correlation is screened as the density is raised for the investigation of exciton Bose Einstein condensate (BEC)~\cite{Griffin} and excitonic insulator (also termed as e-h BCS state)~\cite{Keldysh2}.

The exciton Mott density is defined as the density above which no bound state can exist. In view of the energy diagram, the Mott transition or crossover~\cite{Crossover} is understood to occur when the band edge is lowered due to the band gap renormalization effect and reaches the 1s exciton energy as the e-h pair density increases, thereby resulting in the exciton binding energy vanishing ~\cite{Zimmerman1978, Haug}. In bulk systems, the 1s exciton energy is considered to be almost unchanged due to the cancellation between the band gap renormalization and the reduction of the exciton binding energy. Such a picture explained quantitatively the behavior of 1s exciton resonance, e.g., in a bulk GaAs as observed in the near-infrared optical pump and probe experiments~\cite{Fehrenbach}. However, the continuous reduction of the exciton binding energy associated with the exciton Mott transition has not yet been experimentally verified clearly. Moreover, recent experiments have shown the signature of excitons near the Mott density~\cite{Hayamizu, Kappei, Suzuki1, Shimano} that cannot be accounted for by the above picture. While the exciton Mott transition has been extensively studied as a fundamental problem in quantum many-body theory~\cite{Yoshioka, Kira, Kwong, Manzke, Semkat, Haug, Zimmerman1978, Zimmerman1985}, the experimental investigation remains elusive, particularly on the behavior of the screened Coulomb interaction and the charge carrier properties that are indispensable for understanding the global phase diagram of the photoexcited e-h system in semiconductors. 

Terahertz time-domain spectroscopy (THz-TDS) sheds a new light on this problem.
The recent developments of the optical pump and THz probe (OPTP) spectroscopy~\cite{Kaindl, Suzuki1, Ulbricht} and the ultrafast mid-infrared pump and probe spectroscopy~\cite{Kubouchi} have made it possible to probe the exciton population with a high temporal resolution through the observation of intra-exciton transitions that are typically located in the THz frequency range (1 THz=4.14 meV). Importantly, THz-TDS allows one to determine the complex dielectric function $\epsilon(\omega)$ in the very relevant energy scale of the exciton binding energy, providing direct information on the screening of the Coulomb interaction between electrons and holes. Indeed, the ultrafast dynamics of plasma screening in GaAs has been elucidated by using the OPTP spectroscopy, showing the buildup of a plasmon~\cite{Huber1}. 

In this Letter, we quantitatively study the exciton Mott transition in Si by using the OPTP spectroscopy. The long carrier lifetime of the e-h system in Si~\cite{Hammond} makes it possible to investigate the density-dependent phenomena in quasi-equilibrium condition without it suffering from the carrier lifetime effect. We evaluated the exciton ionization ratio through quantitatively analyzing the dielectric function and conductivity spectra, from which we could determine the Mott density unambiguously. This allows us to confirm that the excitons exist above the Mott density. A Drude-Lorentz model analysis reveals the prominent enhancement of charge carrier scattering rate slightly above the Mott density, rendering the metallic e-h plasma phase highly correlated. Moreover, the loss function spectra ${\rm Im}(-1/\epsilon(\omega))$ that corresponds to the density-density correlation function exhibits a new collective mode of plasmon coupled to the excitonic polarization, which is apparently beyond the picture of standard single-plasmon pole approximation.

To investigate the spectral range covering the exciton binding energy of 14.7 meV in Si, we performed broadband THz spectroscopy as schematically shown in Fig. 1(a). For a light source, we used a Ti:sapphire-regenerative amplifier with the pulse duration of 25 fs, the repetition rate of 1 kHz, and the center wavelength of 800 nm (1.55 eV). The output from the amplifier was divided into three beams for THz generation, THz detection, and for the above-gap optical excitation of e-h pairs, respectively. The THz-probe pulse was generated from a 300 $\mu m$ thick (110)-oriented GaP crystal by the optical rectification of the laser pulse. To detect the THz-probe pulse transmitted after the sample, the electro-optic sampling method was adopted by using a (110)-oriented GaP crystal. The obtained THz spectral range was between 0.5 THz (2 meV) to 6 THz (25 meV). For a sample, we used a high-purity and high-resistivity (10 k$\Omega$cm at room temperature) Si single crystal of (001) surface. The sample was polished to a thickness of 22 $\mu$m, which was thinner than the penetration depth of the optical pump at the measurement temperature (27 $\mu$m at 30 K), and density was distributed almost uniformly in the depth direction. The sample was freely mounted inside a He-gas flow cryostat and kept at temperature higher than 30 K in all the measurements, i.e. above the critical temperature of e-h droplets (EHD) in Si, $T_c$=23 K~\cite{Forchel}, to avoid the formation of EHD.

\begin{figure}[htbp]
\includegraphics[width=80mm]{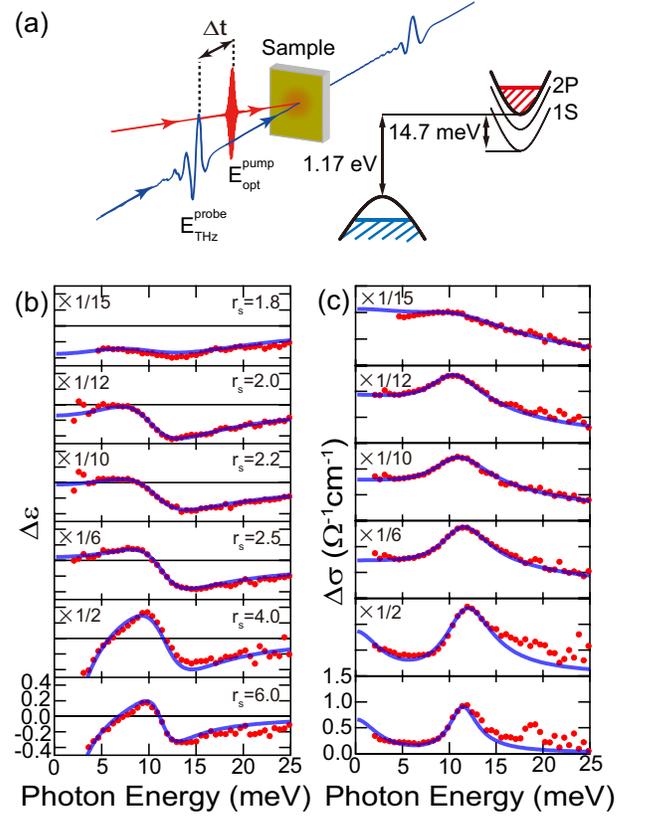}
\caption{(Color online) (a) Schematics of the optical-pump and THz-probe experiment and the band structure of Si. (b) Photo-induced change of dielectric function and (c) optical conductivity at the indicated e-h pair density expressed in terms of $r_s$ parameter (mean interparticles distance normalized by the exciton Bohr radius). The sample temperature is kept at 30 K, and the pump-probe delay time $\Delta t$ is fixed to 4 ns. The solid lines represent Drude-Lorentz fits (see text for details).}
\label{fig1}
\end{figure}

\begin{figure}[htbp]
\includegraphics[width=80mm]{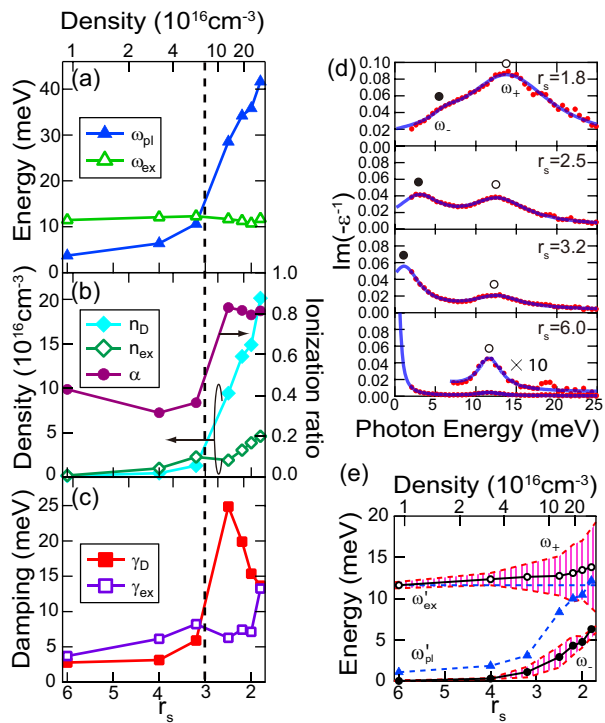}
\caption{(Color online) (a)-(c) The parameters determined from the Drude-Lorentz fits in Fig. 1(b) and (c). (a) The plasma frequency (closed triangles) and the exciton 1s-2p transition energy (open triangles), (b) the density of unbound e-h pairs (closed diamonds), 1s exciton (open diamonds) and the ionization ratio (closed circles), (c) the damping of the Drude term (closed squares) and the Lorentz term (open squares). The vertical dotted line represents the RPA Mott density ($r_s={\text{3.0}}$) at 30 K. (d) The loss function spectra at $T_L$=30 K and $\Delta t=$4 ns at the indicated $r_s$ parameter. The solid curves show the double Lorentz fits. (e) Density dependence of two peaks ($\omega_{+}$ and $\omega_{-}$) identified in the loss function spectra. The shaded area indicates the width of the two modes obtained from the fitting in (d). Dashed lines; the exciton 1s-2p resonance ($\omega^{\prime}_{\text{ex}}$) and the screened plasma frequency ($\omega^{\prime}_{pl}$) determined from (a).}
\label{fig2}
\end{figure}

Figure 1(b) and (c) show the photo-induced change of dielectric function ($\epsilon(\omega)$) and optical conductivity ($\sigma(\omega)$) at the different e-h pair densities expressed in terms of $r_s$ parameter, $r_s=\left(3/4\pi na_{\text{B}}^3 \right)^{1/3}$ where $n$ is the total e-h pair density and $a_{\text{B}}$ is the exciton Bohr radius. $n$ is determined from the excitation pulse energy flux and the absorption coefficient. The sample temperature is kept at 30 K and the delay time $\Delta t$ between the pump pulse and the probe pulse is fixed to 4 ns where the e-h system is considered to reach thermal equilibrium with the lattice system~\cite{Suzuki2}. Note that the Mott density at 30 K calculated from the random phase approximation (RPA) theory is $r_s =3.0$~\cite{Norris}. In the low-density region of $r_s = 6.0$, the dielectric function and the conductivity spectra are dominated by the excitonic 1s-2p resonance around 12 meV with an additional Drude component as indicated by the negative dielectric constant in the low-energy region below 5 meV that is caused by thermally ionized unbound e-h pairs. As e-h pair density increases up to $r_s =2.0$, the sign of dielectric function below 5 meV changes from negative to positive, indicating the suppression of the Drude component. Notably, the excitonic response is sustained even above the Mott density as manifested by the prominent dispersive structure in $\epsilon(\omega)$ as well as by the peak structure around 12 meV in the $\sigma(\omega)$ spectrum. At higher density ($r_s =1.8$), the dielectric function becomes negative in the whole spectral range, indicating the metallic nature of the e-h system, whereas the excitonic structure is still discerned. Strikingly, the excitonic 1s-2p transition energy barely changes as density increases, in contrast to the conventional picture of the exciton Mott transition that predicts the continuous reduction of the exciton binding energy. All these features show the robustness of excitonic e-h pair correlation in the vicinity of Mott density.

To view more quantitatively, we fitted the experimental results with the Drude-Lorentz model, according to which the dielectric function is written as
\begin{equation}
 \epsilon(\omega)=\epsilon_{\text{b}} - \frac{\omega_{\text{pl}}^2}{\omega(\omega + i \gamma_{\text{D}})} - \frac{n_{\text{ex}} e^2}{\epsilon_0 \mu}\frac{1}{\omega^2 - \omega_{\text{ex}}^2 + i\omega \gamma_{\text{ex}}},
\end{equation}
where $\epsilon_{\text{b}}=$11.7 is the background dielectric constant~\cite{Exter}, $\gamma_{\text{D}}$ and $\gamma_{\text{ex}}$ are the damping rate of free carriers (unbound e-h pairs) and 1s excitons, respectively. $\omega_{\text{pl}}=\sqrt{n_{\text{D}}e^2/\epsilon_0 m^{\ast}}$ is the plasma frequency with $n_{\text{D}}$ the unbound e-h pair density, $\epsilon_0$ and $m^\ast=0.16 m_0$ being the vacuum permittivity and optical mass of free carriers in Si~\cite{Riffe}. In the low-density region where the excitons can be viewed as well-defined bound states, $n_{\text{ex}}$ corresponds to the 1s exciton density. $\mu=0.123 m_0$ is the reduced exciton mass~\cite{Nunzio}.

The experimental results of $\epsilon(\omega)$ and $\sigma(\omega)$ are well reproduced by the Drude-Lorentz fits, as shown by the solid lines in Fig. 1(b) and (c). A slight deviation is observed in the high energy side of $\sigma(\omega)$ spectrum at $r_s=$6.0 and 4.0, which can be attributed to the higher lying exciton Rydberg states, which are not taken into account in Eq. (1). This deviation decreases as the e-h pair density increases, plausibly due to the suppression of higher Rydberg states with large Bohr radius in the high-density regime. The parameters determined from the fits are summarized in Fig. 2(a)-(c). Figure 2(a) clearly shows that $\omega_{\text{ex}}$ barely changes across the Mott density($r_s=3$). From the unbound e-h pair density $n_{\text{D}}$ extracted from the plasma frequency, and the exciton density $n_{\text{ex}}$, we can determine the ionization ratio as defined by $\alpha=n_{\text{D}}/(n_{\text{D}}+n_{\text{ex}})$ as plotted in Fig. 2(b). Note that the total density $n_{\text{D}}+n_{\text{ex}}$ obtained from the Drude-Lorentz fits is consistent with the $r_s$ parameter within the accuracy of 10 \% at all densities. As the density is raised, the ionization ratio first decreases toward the Mott density and rapidly increases at the Mott density. Interestingly, the ionization ratio does not reach unity above the Mott density, which originates from the remaining excitonic structure in $\epsilon (\omega)$ and $\sigma (\omega)$ spectra above the Mott density. Such a density dependence of the exciton ionization ratio coincides quantitatively with the theoretical calculation~\cite{Zimmerman1985, Semkat, Stolz} where the existence of a correlation in the scattered state of e-h pairs above the Mott density was demonstrated. Recently, the pair correlation in the e-h plasma was shown to appear as a broad structure near the exciton 1s-2p resonance in the THz absorption spectrum ~\cite{Kira}. At the highest density of $r_s=1.8$, the scattering rate of Lorentz term ($\gamma_X$) is comparable to the transition energy ($\omega_\text{ex}$), and therefore the excitonic structure observed in such high density on the top of the broad Drude-like structure should be better viewed as e-h pair correlation rather than the population of well-defined 1s excitons. However, the damping rate of the Lorentz term slowly increases from below the Mott density (Fig. 2(c)). This behavior indicates that the excitons sustained above the Mott density smoothly change into the unbound but correlated e-h pairs, keeping their resonance intact. In contrast, the ionization ratio shows a steep increase at the Mott density, suggesting that above the Mott density the photo-injected carriers are mostly transferred into free carriers but not to the sustaining excitons. If there exists a spatially inhomogeneous distribution or a phase separation of charge carrier density, excitons may be sustained above the Mott density in the dilute region between the dense carrier region. However, we consider such a scenario to be implausible since, as shown in Fig. 2 (a), the plasma frequency, i.e. the free carrier density, continuously increases across the Mott density without exhibiting the signature of first order phase transition that results in the phase separation.

Interestingly, the Drude scattering term is found to increase drastically just above the Mott density and then decreases as the density is raised as shown in Fig. 2(c). This prominent enhancement of the Drude scattering rate accounts for the suppression of negative contribution in the dielectric function at $r_s=$2.5 and 2.2, giving rise to the positive dielectric constant in the low-energy region below 5 meV. Such a nonmonotonic behavior of Drude scattering term is not accounted for by electron-hole scattering, which should be approximately linear with density. Instead, the main contribution to the enhancement of Drude scattering rate is attributed to the interaction between the charge carriers and the excitons, which is further confirmed by the temperature dependence as shown later (Fig. 3(c)). Although the e-h system at the measured temperature of 30 K does not degenerate since $k_{\text{B}}T=$ 2.6 meV is larger than the $E_F\sim $ 1 meV at $r_s$=2.5 for both electrons and holes, the observed scattering rate of 25 meV is one order of magnitude larger than the kinetic energy of free carriers, suggesting the bad metal nature of charge carriers in this density region.

Having seen that the exciton is sustained in the metallic e-h plasma, we plot in Fig. 2(d) the loss function spectra ${\rm Im}(-1/\epsilon(\omega))$. Note that the peak of Im($-1/\epsilon(\omega)$) indicates the longitudinal collective modes of the system. In the low-density region, two separate peaks are clearly identified; the sharp peak at the lower-energy side corresponds to the (${\bf q}=0$) plasmons and the higher one corresponds to the longitudinal mode associated with the exciton 1s-2p transition. As e-h pair density increases, the plasmon resonance shifts to the high-energy side but does not intersect with the exciton branch. The anti-crossing behavior is more clearly seen in Fig. 2(e), where we plot the density($r_s$)-dependence of the peak positions determined from a phenomenological double-Lorentzian fits (solid lines in Fig. 2(d)). The hatched area indicates the FWHM of each mode determined from the Lorentzian fits. We also plot in Fig. 2(e) the bare exciton 1s-2p resonance and the screened plasma frequency $\omega_{\text{pl}}^\prime=\omega_{\text{pl}}/\sqrt{\epsilon_b}$ obtained from Fig. 1(b) (dashed lines). In the density region where the level anti-crossing occurs and therefore the plasmon-exciton coupling is pronounced, the plasmon rapidly broadens. This broadening of the plasmon mode is ascribed to the existence of the exciton. While the plasmon damping caused by the single-particle excitations is restricted at ${\bf q}=0$, here the plasmon damping is caused by the presence of the exciton to which the plasmon couples. Such a coupling between plasmons and other bosonic excitations is generic when their resonances come closer, as exemplified by the plasmon-LO phonon coupling~\cite{Mooradian}. In the present case, however, the coupling is purely electronic, suggesting the coupled nature of long-wavelength charge density fluctuation with the excitonic e-h pair correlation. 

Figure 3 (a) presents the temperature dependence of the loss function spectra ${\rm Im}(-1/\epsilon(\omega))$ under the excitation condition of $r_s = 3.1$ and at the pump-probe delay of 4 ns. The loss function spectra at several temperatures are also shown in Fig. 3(b). At high temperature, the peak at 5 meV is assigned to the pure plasmon mode. As the temperature is lowered to 100 K, the peak gradually sharpens due to the reduction of scattering rate with phonons. Below 100 K, an additional peak emerges around 12 meV, although broadened, which is assigned to the 1s-2p excitonic transition. At lower temperature, the growth of excitonic structure is discerned and correspondingly the red shift of the plasmon resonance is observed. This feature indicates the spectral weight transfer from free electrons and holes to excitons. Concomitantly, the plasmon peak rapidly broadens as shown in Fig. 3(b). Below 40 K, two peaks are clearly distinguished, while these resonances should be viewed as the coupled modes of plasmons and excitons. Figure 3(c) plots the temperature dependence of the Drude scattering rate determined from the Drude-Lorentz fits at different density of $r_s=6.0$ and $r_s=3.1$. As the temperature is raised, the scattering rate first decreases and then exhibits an upturn. This nonmonotonic temperature dependence becomes more pronounced at $r_s=3.1$ near the Mott density. It is interesting to note that recently a similar behavior has been reported in the low-energy optical response of Bi in the vicinity of semimetal to semiconductor transition, where the pronounced damping of the Drude scattering rate was interpreted as evidence of the coupling of plasmons to charge carriers, namely the plasmarons~\cite{Tediosi, Armitage}. In the present case of the photoexcited e-h system, the nonmonotonic temperature and density dependence of the Drude scattering rate, which is pronounced at the proximity of Mott density, originates from the exciton that emerges below 60 K.

\begin{figure}[htbp]
\includegraphics[width=85mm]{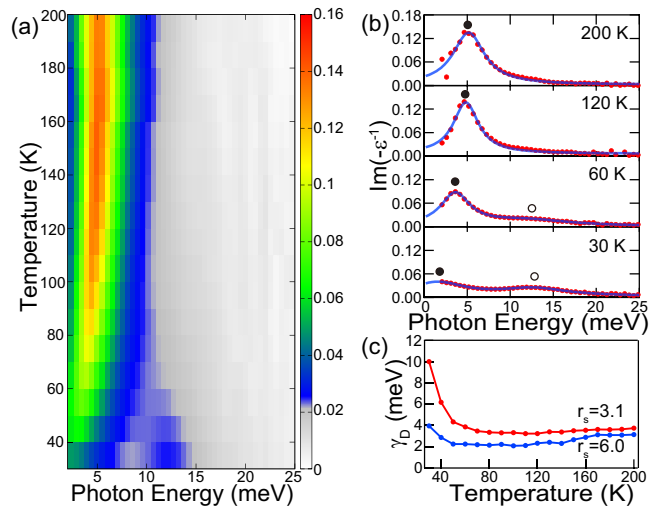}
\caption{(Color online) (a)Temperature dependence of the loss function spectra at $\Delta t=$4 ns and $r_s=3.1$. (b) Loss function spectra at indicated temperature. Solid curves show the fitting with a double Lorentz model. The peak positions are indicated by the closed and open circles. (c) Temperature dependence of the Drude scattering rate (Eq.(1)) for $r_s=3.1$ and $r_s=6$.}
\label{fig1}
\end{figure}

In summary, we have investigated the exciton Mott transition in Si by using OPTP spectroscopy. Moreover, THz spectroscopy enabled us to evaluate quantitatively the exciton ionization ratio from which the Mott density is unambiguously determined. The energy position of the excitonic 1s-2p transition barely changes as the density increases and is sustained in the metallic phase above the Mott density. The Drude scattering rate of charge carriers is largely pronounced near above the Mott density, indicating the emergence of highly correlated metallic phase due to the existence of nonvanishing excitons. Concomitantly, the loss function spectrum exhibits coupled behavior of plasmons and excitons near the Mott density, giving rise to a new collective mode of charge density excitation. Since THz spectroscopy directly provides the information of e-h pair correlation through the spectrum of the dielectric function in the very relevant energy scale, it would be useful to explore the possibility of exciton BEC and e-h BCS, and to investigate further IMT in a wide range of materials.

We wish to acknowledge helpful discussions with T. M. Rice. This work was in part supported by Grant-in-Aid for Scientific Research (Grants No. 22244036, No. 23104705, No. 20110005) and by the Photon Frontier Network Program from MEXT, Japan.

\end{document}